\documentstyle[12pt]{article}
\begin{document}
\tolerance=5000
\def\be{\begin{equation}}
\def\ee{\end{equation}}
\def\bea{\begin{eqnarray}}
\def\eea{\end{eqnarray}}
\def\beas{\begin{eqnarray*}}
\def\eeas{\end{eqnarray*}}
\def\nn{\nonumber \\}
\def\cF{{\cal F}}
\def\det{{\rm det\,}}
\def\Tr{{\rm Tr\,}}
\def\e{{\rm e}}
\def\etal{{\it et al.}}
\def\erp2{{\rm e}^{2\rho}}
\def\erm2{{\rm e}^{-2\rho}}
\def\er4{{\rm e}^{4\rho}}
\def\etal{{\it et al.}}

\  \hfill 
\begin{minipage}{3.5cm}
OCHA-PP-125 \\
NDA-FP-51 \\
October 1998 \\
\end{minipage}

\ 

\vfill

\begin{center}

{\large\bf D0-brane description of the charged black hole}

\vfill

{\sc Yuriko KATO}\footnote{e-mail : g9740507@edu.cc.ocha.ac.jp} \\
{\sc Shin'ichi NOJIRI$^{\clubsuit}$}\footnote{
e-mail : nojiri@cc.nda.ac.jp} and 
{\sc Akio SUGAMOTO}\footnote{
e-mail : sugamoto@phys.ocha.ac.jp}

\vfill

{\sl Department of Physics, 
Ochanomizu University \\
Otsuka, Bunkyou-ku Tokyo 112, JAPAN}

\ 

{\sl $\clubsuit$ 
Department of Mathematics and Physics \\
National Defence Academy, 
Hashirimizu Yokosuka 239, JAPAN}

\vfill

{\bf ABSTRACT}

\end{center}

The charged black hole is considered from the viewpoint of D0-brane in 
the Matrix theory. It can be obtained from the Kaluza-Klein mechanism by 
boosting the Schwarzschild black hole in a circle, which is the 
compactified one dimensional space. Especially, how the extremal limit 
is realized by the Boltzmann gas of D0-brane, has been shown. In the 
course of our discussion, the Virial theorem for the statistical average 
plays an important role.

\ 

\noindent
PACS: 

\newpage

It is known that the several properties of the Schwarzschild black holes 
can be well described by the Boltzmann gas of the D0-brane \cite{BFKS,Li}, 
the parton of the Matrix theory \cite{BFSS}. Recently Ohta and Zhou 
\cite{Ohta} have calculated the partition function of the Matrix theory in 
the D0-brane background. They have derived the mass, entropy, temperature 
etc. of this system and shown that the D0-brane background can be 
interpreted as 11D Schwarzschild black hole. 

Since D0-branes interact with each other very weakly, the most of the 
properties of the D0-brane background can be derived from the description 
of the free Boltzmann gas. The information of the interaction can be 
included by using the Virial theorem. 

In this letter, we treat the charged black hole in the Matrix theory, which 
can be obtained from the Kaluza-Klein mechanism by boosting the 
Schwarzschild black hole in a circle, which is the compactified 
one dimensional space. Especially, we show how the extremal limit is 
described by the Boltzmann gas of D0-brane. In \cite{Ohta}, the oscillation 
modes in the Matrix theory is integrated out by using the Euclidean 
path integral method. In this letter, we give the more statistical method for 
the treatment of the zero modes, which reproduces the same results of 
the Schwarzschild black hole in \cite{Ohta} derived from the consistency, 
and gives the new results on the charged black hole. 

We start with the free part of the M-theory effective Lagrangian in 
\cite{Ohta}, which is given by integrating oscillating modes in a given 
background $\bar X^i$:
\be
\label{i}
{\cal L}^{\rm free}={\rm Tr}{1 \over 2R}(\dot{\bar X^i})^2 \ ,
\ee
where $R$ is the radius of the light-like compactification from 11D to 10D. 
Then the canonical momentum $P^i$ conjugate to $\bar X^i$ reads
\be
\label{ii}
P^i={1 \over R}\dot{\bar X^i}\ .
\ee
As a background $\bar X^i$, we consider the following one in which $N$ 
D0-branes are moving freely :
\be
\label{iii}
\bar X^i={\rm diag.}(v_1^it+b_1^i, v_2^it+b_2^i, \cdots, 
v_N^it+b_N^i)\ .
\ee
If the black hole is a statistical system in which D0-branes are moving 
inside the sphere of the radius $R_s$ at temperature $\beta^{-1}$ in the 
light-cone frame, we can define the partition function of the black hole 
as follows:\footnote{The explicit integrations for the zero modes $v_n^1$ 
and $b_n^i$ have been not given in \cite{Ohta}.}
\bea
\label{iv}
Z(\beta, N)&=&\int d\bar X^i 
\e^{-\int_{-{\beta \over 2}}^{\beta \over 2}{\cal L}^{\rm free}}
\nn
&\sim&{1 \over R^{9N}}\int_{-\infty}^\infty \prod dv_n^i
\int_{-{R_s \over 2}}^{R_s \over 2} \prod db_n^i
\e^{-{\beta \over 2R}(v_n^i)^2} \nn
&=&\left({2\pi R_s^2 \over \beta R }\right)^{9N \over 2}\ .
\eea
Note that, from (\ref{ii}), the phase space volume for a pair of $n$-th 
coordinate and its conjugate momentum of the background D0-brane is given 
by (we choose $\hbar=1$) 
\be
\label{iva}
dx_n^idp_n^i={dv_n^i db_n^i \over R}\ .
\ee

By using (\ref{iv}), we find 
\bea
\label{v}
<v_n^2>&=&-{2R \over N}{\partial \over \partial \beta}\ln 
Z(\beta, N) \nn
&=&{9R \over \beta}\ .
\eea
Since $\beta$ is the Euclidean time interval, the typical magnitude of 
the velocity of D0-brane moving within the black hole having the radius 
$R_s$ is given by
\be
\label{vi}
<v_n^2>\sim{R_s^2 \over \beta^2}\ .
\ee
By comparing (\ref{v}) and (\ref{vi}), we find
\be
\label{vii}
\beta\sim{R_s^2 \over R}\ ,
\ee
which corresponds to Eq.(2.9) in ref.\cite{Ohta}. By using (\ref{iv}), we 
find the Helmholz free energy $F(\beta, N)$ 
\bea
\label{viii}
F(\beta, N)&=&-{1 \over \beta}\ln Z(\beta,N) \nn
&=& -{9N \over 2\beta}\ln\left({2\pi R_s^2 \over \beta R }\right) \ ,
\eea
and the entropy $S$
\bea
\label{ix}
S(\beta, N)&=&\beta^2 {\partial F \over \partial \beta} \nn
&=&{9N \over 2}\left\{\ln
\left({2\pi R_s^2 \over \beta R }\right) +1\right\}\ .
\eea
 From (\ref{vii}), we obtain
\be
\label{x}
S(\beta, N)\sim N
\ee
which corresponds to (3.11) in \cite{Ohta}. 
The expectation value of the energy $E$ is also given by 
\bea
\label{xb}
E&=&-{\partial \over \partial \beta}\ln 
Z(\beta, N) \nn
&=&{9N \over 2\beta}\ .
\eea

We now use the expression (3.1) in \cite{Ohta} for $E$ 
\bea
\label{xi}
E&=&\sum_{n=1}^N{v_n^2 \over 2R}
-{15G_{11} \over 16R^3}\sum_{n<m}^N{v_{nm}^4 \over 
\left(b_{nm}^2 + {v_{nm}^2\beta^2 \over 4}\right)^{7 \over 2}} \\
v_{nm}^i&\equiv&v_n^i - v_m^i \nn
b_{nm}^i&\equiv&b_n^i - b_m^i \ . \nonumber 
\eea
This is the effective energy for a given background configuration, including 
the fluctuations arround it, but the statistical averaging over the 
background configurations is further required. Therefore we apply the 
Virial theorem to (\ref{xi}) by regarding $E$ in (\ref{xi}) as 
the Hamiltonian and assume
\be
\label{xii}
v_n^2\sim v_{nm}^2\sim {R_s^2 \over \beta^2}\sim {R \over \beta}
\ ,\ \ b_{nm}^2 \sim R_s^2
\ee
as in (\ref{vi}). Then we find 
\be
\label{xiii}
R_s\sim (G_{11}N)^{1 \over 9}\ ,
\ee
The derivation of which will be clarified in 
the following\footnote{Essentially we assume that the expectation values 
of the first and the second terms are the same magnitude with each other.}.
By using (\ref{vii}), (\ref{xb}) and (\ref{xiii}), we find also 
\be
\label{xiv}
E\sim{\left(G_{11}^{-{1 \over 9}}N^{8 \over 9}\right)^2 R \over N}\ .
\ee
The relations (\ref{xiii}) and (\ref{xiv}) correspond to (3.9) in \cite{Ohta}.

Usually the Virial theorem is given by the average with respect to time. 
In the language of the statistical average, the theorem can be proved in the 
following way: Here we only consider the case in that Hamiltonian $H$ is given 
by a pair of one coordinate $q$ and its conjugate momentum $p$:
\be
\label{VVi}
H=H(q,p)\ ,
\ee
since the generalization to the case with several coordinates and 
their conjugate momenta is straightfoward. We can observe that the 
expectation value of $p{\partial H \over \partial p}-q{\partial H 
\over \partial q}$ vanishes, since it is the total derivative in the 
phase space
\bea
\label{VVii}
&& \left\langle 
p{\partial H \over \partial p}-q{\partial H \over \partial q} 
\right\rangle \nn
&& = {1 \over Z(\beta)}\int dqdp \left(
p{\partial H \over \partial p}-q{\partial H \over \partial q} 
\right) \e^{-\beta H(q,p)} \nn
&& = {1 \over Z(\beta)}\int dqdp \left\{{\partial \over \partial p}
\left(p\e^{-\beta H(q,p)} \right) - {\partial \over \partial q}
\left(q\e^{-\beta H(q,p)} \right) \right\} \nn
&& = 0\ , 
\eea
where
\be
\label{VViii}
Z(\beta)\equiv \int dqdp\, \e^{-\beta H(q,p)} \ .
\ee
The first term $T=p{\partial H \over \partial p}$ corresponds to the 
kinetic energy and the second term $V=q{\partial H \over \partial q}$ 
to the so-called ``virial'', so that we find the Virial theorem as 
\be
\label{VViv}
\langle T \rangle = \langle V \rangle \ .
\ee
By applying the above theorem to (\ref{xi}) and using (\ref{xii}), we 
find (\ref{xiii}). 

Now we extend the above analysis to the case of charged black hole. 
The charged black hole can be obtained by the Kaluza-Klein mechanism
\cite{HHS}. We impose a periodic boundary condition on one dimension in 
11 dimensional Schwarzschild black hole spacetime and compactify it to 
a circle. After that, we boost the black hole solution in that one 
dimensional direction. Then the total momentum in the compactified space 
can be regarded as the total charge of the black hole.

Let $X^1$ be the coordinate in the compactified space. We now project 
the partition function in (\ref{iv}) to the sector in which total momentum 
conjugate to $X^1$ is ${V \over R}$, by inserting a delta function. 
By using the following expression of a delta function
\be
\label{xv}
\delta\left(\sum_{n-1}^N v_n^1 - V\right) ={1 \over 2\pi}
\int_{-\infty}^\infty dk \e^{ik\left(\sum_{n=1}^N v_n^1 - V\right)}\ ,
\ee
we obtain 
\bea
\label{xvi}
&&{1 \over 2\pi}\int_{-\infty}^\infty \prod_{n=1}^N dv_n^1
\int_{-\infty}^\infty dk \e^{-{\beta \over 2R}
\sum_{n=1}^N (v_n^1)^2+ik \left(\sum_{n=1}^N v_n^1 - V\right)} \nn
&& = {1 \over 2\pi}\int_{-\infty}^\infty \prod_{n=1}^N dv_n^1
\int_{-\infty}^\infty dk \e^{-{\beta \over 2R}
\sum_{n=1}^N \left(v_n^1 - {ikR \over \beta}\right)^2
-{NR \over 2\beta}\left(k+{i\beta V \over NR}\right)^2
-{\beta V^2 \over 2NR}} \nn
&& =\left({\beta \over 2\pi NR}\right)^{1 \over 2}
\e^{-{\beta V^2 \over 2NR}} \int_{-\infty}^\infty \prod_{n=1}^N dv_n^1
\e^{-{\beta \over 2R} \sum_{n=1}^N (v_n^1)^2}
\eea
and we find that the projected partition function is given by 
\be
\label{xvii} 
Z(\beta,N,V)=\left({2\pi R_s^2 \over \beta R }\right)^{9N \over 2}\left(
{\beta \over 2\pi NR}\right)^{1 \over 2}\e^{-{\beta V^2 \over 2NR}} \ .
\ee
In a similar way to derive (\ref{v}), from the partition function in 
(\ref{xvii}), we obtain
\be
\label{xviii}
<v_n^2>=\left(9R -{R \over N}\right){1 \over \beta}+{V^2 \over N^2}\ .
\ee
By assuming $<v_n^2>\sim{R_s^2 \over \beta^2}$ as in (\ref{vi}), 
we can solve Eq.(\ref{xviii}) with respect to $\beta$ and obtain 
\be
\label{xix}
{1 \over \beta}={9R \pm \sqrt{\left( 9R \right)^2
+ {4R_s^2 V^2 \over N^2}} \over 2R_s^2}\ .
\ee
Here we also assume $N\gg 1$. In (\ref{xix}), the negative sign cannot 
be physically accepted since it gives the negative temperature. The 
expression of the entropy $S$ in (\ref{ix}) does not be changed. The 
expectation value of the energy is given by
\be
\label{xx}
E={9N-1 \over 2\beta}+{V^2 \over 2NR}\ .
\ee

Since $E$ is related with the black hole mass $M$ \cite{BFKS} as follows
\be
\label{xxb}
E={M^2 R \over 2N}, 
\ee
we find 
\be
\label{xxi}
M^2 = {(9N-1)N \over R\beta}+{V^2 \over R^2}\ .
\ee
Furthermore by using the relation between the light-cone temperature 
$\beta^{-1}$ and the rest frame temperature $T$ \cite{Ohta},
\be
\label{xxii}
{1 \over \beta}={MR \over 2N}T\ , 
\ee
we find 
\be
\label{xxiii}
T={2\left(M^2 - {V^2 \over R^2}\right) \over (9N-1)M}\ .
\ee
Remarkably Eq.(\ref{xxiii}) tells that the temperature vanishes when 
\be
\label{xxxiv}
M=Q\equiv |{V| \over R}\ .
\ee
Therefore Eq.(\ref{xxxiv}) is nothing but the condition of the 
extremal limit of the black hole.

In summary, we have derived the mass, temperature etc. of the charged 
balck hole by using the D0-brane description in the Matrix theory. 
Especially, we have found how the exremal limit is described.

\end{document}